\documentclass[conference]{IEEEtran}
\IEEEoverridecommandlockouts
\usepackage{cite}
\usepackage{amsmath,amssymb,amsfonts}
\usepackage{graphicx}
\usepackage{textcomp}
\usepackage[table]{xcolor}
\usepackage[hidelinks]{hyperref}
\usepackage{mdframed}
\usepackage{tabularray}
\usepackage{tikz}
\newcommand*\circled[1]{\tikz[baseline=(char.base)]{
            \node[shape=circle,draw,fill=black,text=white,inner sep=1pt] (char) {#1};}}
\usepackage{multirow}
\def\BibTeX{{\rm B\kern-.05em{\sc i\kern-.025em b}\kern-.08em
    T\kern-.1667em\lower.7ex\hbox{E}\kern-.125emX}}

\makeatletter
\def\blfootnote{\xdef\@thefnmark{}\@footnotetext}
\makeatother

\usepackage{algorithm,algpseudocode}
\newcommand{\algmargin}{\the\ALG@thistlm}
\makeatother
\algnewcommand{\parState}[1]{\State%
    \parbox[t]{\dimexpr\linewidth-\algmargin}{\strut\hangindent=\algorithmicindent \hangafter=1 #1\strut}}

\usepackage{wrapfig}
\usepackage{enumitem}
\usepackage[switch]{lineno}
\begin{document}

\title{SAGE: A Context-Aware Approach for Mining Privacy Requirements Relevant Reviews from Mental Health Apps
}

\author{
    \IEEEauthorblockN{Aakash Sorathiya, Gouri Ginde}
    \IEEEauthorblockA{\textit{Department of Electrical and Software Engineering} \\
        \textit{University of Calgary}\\
        Calgary, Canada \\
        \{aakash.sorathiya, gouri.ginde\}@ucalgary.ca}
}
\maketitle

\begin{abstract}
Mental health (MH) apps often require sensitive user data to customize services for mental wellness needs. However, such data collection practices in some MH apps raise significant privacy concerns for users. These concerns are often mentioned in app reviews, but other feedback categories, such as reliability and usability, tend to take precedence. This poses a significant challenge in automatically identifying privacy requirements-relevant reviews (privacy reviews) that can be utilized to extract privacy requirements and address users' privacy concerns. Thus, this study introduces SAGE, a context-aware approach to automatically mining privacy reviews from MH apps using Natural Language Inference (NLI) with MH domain-specific privacy hypotheses (provides domain-specific context awareness) and a GPT model (eliminates the need for fine-tuning). The quantitative evaluation of SAGE on a dataset of 204K app reviews achieved an F1 score of 0.85 without any fine-tuning, outperforming the fine-tuned baseline classifiers BERT and T5. Furthermore, SAGE extracted 748 privacy reviews previously overlooked by keyword-based methods, demonstrating its effectiveness through qualitative evaluation. These reviews can later be refined into actionable privacy requirement artifacts.
\end{abstract}


\section{Introduction}
\textbf{Context.} In 2024, 23.08\% of U.S. adults experienced untreated Mental Health (MH) challenges, affecting nearly 60 million due to access barriers and high costs \cite{reinert2024state}. Recent advancements in mobile technology have produced affordable mobile apps to improve access to MH services \cite{bauer2020smartphones}. The Google Play and Apple App stores host over 20,000 MH apps, with some exceeding 10 million downloads and consistent usage \cite{eagle2022money}. However, in several nations such as the US and Canada, MH apps remain predominantly unregulated, classified as addressing `well-being' or `lifestyle' \cite{torous2018dichotomies} and are considered `minimal risk' \cite{mcniel2018current}. This regulatory oversight of MH apps has raised serious concerns about privacy, security, and usability \cite{torous2017needed}. Thus, app developers are increasingly required to assess and scrutinize user feedback for privacy concerns, such as data collection practices, to ensure market viability \cite{khalid2014mobile}.

\textbf{Problem.} Users provide feedback through textual app reviews \cite{aamir2023government, ebrahimi2021mobile}, widely used for various requirements engineering (RE) tasks \cite{dkabrowski2023mining, dkabrowski2022analysing}, including requirements extraction from privacy reviews \cite{sorathiya2024towards}. Several studies that analyzed app reviews for privacy concerns investigated mainly users’ perspectives on privacy concerns within mobile apps \cite{mcilroy2016analyzing, besmer2020investigating, nema2022analyzing, harkous2022hark, ebrahimi2022unsupervised, nguyen2019short, mukherjee2020empirical, dong2022privacy}. Nguyen et al. \cite{nguyen2019short} used a traditional machine learning model to classify privacy reviews and investigated how app reviews impact privacy app updates. Nema et al. \cite{nema2022analyzing} and Ebrahimi et al. \cite{ebrahimi2022unsupervised} emphasized summarizing app reviews to identify common privacy concerns. These studies primarily rely on keyword or regular expression-based sampling techniques, which may lack contextual information and restrict privacy concerns to specific keywords. 

For instance, in the following review, \textit{``My husband and I need counseling. We thought this was a great idea to give this a try. Unfortunately we were wrong! I spoke with my "consultant" Jennifer K. She was very chatty and seemed helpful. Then she provided the link to pay. I paid 189\$ for 1 month of couples therapy. She then provided me a link for my husband to join us in the consult private room. The first link did not work at all. The second one she provided took him to a different consultant Anne..."}, ``private" keyword indicates a private consultation room; however, it was considered as a privacy-indicative keyword in the previous study \cite{ebrahimi2022unsupervised}. Consequently, identifying privacy reviews is significantly dependent on contextual interpretation; thus, merely searching for related keywords might not be an effective approach. Furthermore, these studies are based on the supervised classification of app reviews that require manual effort and can be time-consuming.

\textbf{Solution. } We propose SAGE\footnote{SAGE is an English noun meaning ``a person of deep wisdom or learning."}, a novel context-aware approach that combines Natural Language Inference (NLI) and a GPT large language model (LLM) to extract privacy reviews (privacy requirements-relevant reviews from which we can extract privacy requirements to address users' privacy concerns). SAGE utilizes NLI with MH domain-specific privacy hypotheses, which provide domain-specific context awareness, to identify potential privacy reviews. It then employs a GPT model, specifically a GPT-4o-mini version of OpenAI's closed-source LLMs, which only utilizes zero-shot learning, to extract actual privacy reviews from the set of potential reviews.

\textbf{Rationale.} NLI assesses the relationship between a premise and a hypothesis \cite{fazelnia2024lessons}. It identifies whether the premise entails, contradicts, or is neutral to the hypothesis. NLI has demonstrated effectiveness in tasks such as question answering \cite{demszky2018transforming} and semantic search \cite{stasaski2022semantic}. It offers a more refined and linguistically consistent evaluation of text relationships, surpassing keyword-based methods by delivering context-aware results \cite{fazelnia2024lessons}. Harkous et al. \cite{harkous2022hark} utilized NLI in the privacy domain, but they rely on generic privacy hypotheses, overlooking that user privacy concerns are domain-dependent \cite{ebrahimi2022unsupervised}. Therefore, we propose MH domain-specific privacy hypotheses to enhance domain-specific context awareness in NLI. GPT is a generative pre-trained transformer model commonly employed as a zero-shot text classifier, negating the need for fine-tuning with labeled data in low-data scenarios \cite{chae2023large}.

\textbf{Contributions.} The main contributions of our study are:
\begin{itemize} [leftmargin=*]
    \item Proposed a novel, scalable, and reproducible hybrid automation framework (SAGE) that integrates domain-specific NLI with a zero-shot GPT-based classifier to automatically identify privacy requirements-relevant user feedback without requiring any fine-tuning or labeled training data.
    \item Developed and proposed a domain-adaptive NLI strategy that operationalizes mental health-specific privacy concepts into 17 structured hypotheses, enabling context-aware filtering of candidate reviews and improving precision over generic hypotheses and keyword-based approaches. 
    \item Provided evidence through a comprehensive empirical evaluation using a publicly available dataset \cite{ebrahimi2021mobile} as our evaluation benchmark (across 204K app reviews), demonstrating how SAGE achieves competitive performance (F1 = 0.85) compared to fine-tuned supervised models (e.g., BERT, T5), while significantly reducing manual annotation effort (Cohen's $k$ = 0.71 with human labels).
    \item Discovered previously undetected 748 privacy reviews, which can be refined into actionable requirements artifacts.
    \item  To support reproducibility and future research in automated requirements engineering (RE) and ethical software analysis, we have made our code open source\footnote{\url{https://doi.org/10.5281/zenodo.15650939}}, including newly extracted privacy reviews and domain-specific hypotheses.
\end{itemize} 

\textbf{Structure. }Section \ref{prem} outlines preliminaries and Section \ref{method} presents our approach. Sections \ref{evaluation} and \ref{results} present the evaluation design and results, respectively. Section \ref{discuss} discusses the implications of our findings, and Section \ref{threats} presents validity threats. Finally, Section \ref{brw} reviews related work and Section \ref{conclude} offers concluding remarks.

\section{Preliminaries} \label{prem}


\noindent\textbf{Natural Language Inference (NLI):}
NLI pertains to the problem of determining whether a hypothesis can logically be derived from a specified premise, i.e., an NLI model identifies whether a hypothesis is true (i.e., \textit{entailment}), false (i.e., \textit{contradiction}), or undetermined (i.e., \textit{neutral}) for a given premise \cite{maccartney2009extended}. For instance, consider a premise stating, ``…collecting all of my Facebook data, just stole my identity…". A hypothesis - ``too much personal data is collected" - would be assigned an \textit{entailment} label. Conversely, a hypothesis - ``user likes that data privacy is provided" - would be designated a \textit{contradiction} label, and a hypothesis - ``app has a good interface" would be assigned a \textit{neutral} label. In our study, the premises are app reviews, and the domain-specific hypotheses are manually derived from the privacy concepts (structured categories of privacy violations) in the MH domain. 

This NLI-based methodology mitigates the dependency on specific keywords due to the extensive linguistic variability present in the premises associated with the hypotheses \cite{harkous2022hark}. For example, the below reviews received an \textit{entailment} label for the hypothesis ``User is not aware of how and why their data is being collected, processed, stored, and shared.":

\begin{itemize}
    \item ``Don't bait people in to take their information and sell it and add them to your mailing list then force a paywall to use the app. Free to download pay to use, 1 star no recommendation due to lack of transparency." (P(\textit{entailment})=0.76)
    \item ``This app has 6 data trackers. Don't trust any app with your "wellbeing" that is sending your behavior data to multiple third parties." (P(\textit{entailment})=0.87)
\end{itemize}

Note that no review has any words in common with the hypotheses, but both of them discuss the concern related to data collection and sharing. Here, P(\textit{entailment}) denotes the probability of the \textit{entailment} label and is referred to as \textit{entailment\_score}. We use these scores to filter out the potential reviews based on the defined heuristics.


\begin{figure*}[htpb]
    \centering
    \includegraphics[scale=0.32]{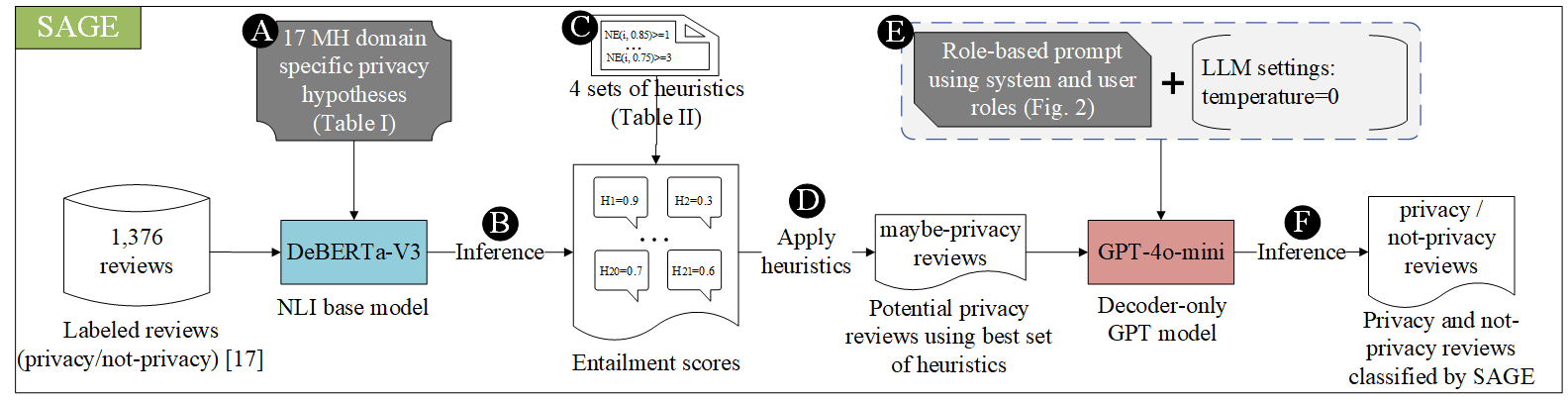}
    \caption{SAGE design process. At the core, we have domain-specific NLI and a GPT model. NLI with MH domain-specific hypotheses is used to extract potential privacy reviews, and then the GPT model is used to classify privacy/not-privacy reviews from the set of potential reviews. At the NLI step, we perform experiments with 4 sets of heuristics to identify the best one and select the best set of potential privacy reviews.}
    \label{fig:design}
\end{figure*}

\section{Approach} \label{method}
Figure \ref{fig:design} provides an overview of the process used to design SAGE. SAGE has two components: NLI inference (to extract potential privacy reviews using MH domain-specific privacy hypotheses) and LLM inference (to extract privacy reviews from a set of potential privacy reviews using the GPT model). First, we derived domain-specific privacy hypotheses using MH domain privacy concepts to perform NLI inference and extracted the \textit{entailment} relation between hypotheses and app reviews. Then, we defined the corresponding heuristics to select potential privacy reviews. Finally, we designed the prompt for the GPT model and performed LLM inference to classify privacy reviews from a set of potential privacy reviews. To perform the quantitative and qualitative evaluation of SAGE, we defined five RQs and used the dataset provided by Ebrahimi et al. \cite{ebrahimi2022unsupervised}. We also manually validated the data extracted by SAGE as part of the qualitative evaluation.

\textbf{\underline{A) NLI Inference}} 
We leverage the NLI model to identify potential privacy reviews discussing certain privacy concepts in the MH domain. Hence, we first defined 17 domain-specific privacy hypotheses (step \circled{A}) and utilized the ground truth dataset from Ebrahimi et al. \cite{ebrahimi2022unsupervised} to perform the inference operation using the NLI model (step \circled{B}). At this stage, we performed 1,376 (number of app reviews) * 17 (privacy hypotheses) = 23,392 inference operations. Then, we defined four sets of heuristics (step \circled{C}) and applied them to the \textit{entailment\_scores} (step \circled{D}) received from the inference operation. We finally selected the best set of heuristics that yielded the best set of potential privacy reviews. 

\textit{Deriving MH domain-specific privacy hypotheses.}
The domain-specific privacy hypotheses were constructed manually based on the privacy concepts of the MH domain provided by Iwaya et al. \cite{iwaya2023privacy} in their exploration of the development of MH apps. Following the method provided by Fazelnia et al. \cite{fazelnia2024lessons}, which is to convert privacy concepts into detailed sentences, we developed one or more hypotheses for each concept. We converted the privacy concepts into hypotheses representing corresponding privacy violations and obtained entailment relations for app reviews. For instance, for the ``\textit{Non-repudiation}" concept, we defined two hypotheses: ``\textit{User cannot deny having performed certain actions within the app}" and ``\textit{User is concerned about the permanent storage of their mental health history}" representing the violation of ``\textit{Non-repudiation}". Following this approach, we defined 17 domain-specific privacy hypotheses as shown in Table \ref{tab:dhypo}.
\begin{table}[h]
    \centering
    \caption{MH domain-specific privacy concepts \cite{iwaya2023privacy} and associated hypotheses. Each privacy concept and corresponding hypotheses represent a privacy violation in the MH domain.}
    \label{tab:dhypo}
    \begin{tabular}{p{1.3cm} p{6.5cm}}
         \multicolumn{2}{l}{\textbf{Concepts from Iwaya et al’s Taxonomy \cite{iwaya2023privacy}}} \\
        \hline
        \textbf{Privacy Concept} & \textbf{Hypotheses} \\
        \hline 
        \multirow{1}{1.5cm}{Linkability} & 1. Mental health data is linked across different services. \\
        & 2. Online activities across various mental health apps can be connected. \\
        & 3. Personal information about users' mental health is collected from external sources. \\ \hline
        \multirow{1}{1.5cm}{Identifiability} & 4. Anonymized mental health data is used to re-identify the user. \\
        & 5. Unique patterns in a user’s psychological data lead to personal identification. \\ \hline
        \multirow{1}{1.5cm}{Non-repudiation} & 6. User cannot deny having performed certain actions within the app. \\
        & 7. User is concerned about the permanent storage of their mental health history. \\ \hline
        \multirow{1}{1.5cm}{Detectability} & 8. User is concerned about others detecting their use of sensitive mental health services. \\
        & 9. Users’ participation in mental health apps is discovered from anonymized usage data. \\ \hline
        \multirow{1}{1.5cm}{Disclosure of Information} & 10. Users’ device communication patterns reveal private information about their mental health conditions. \\
        & 11. Mental health data intercepted during transmission. \\
        & 12. Mental health app exposes a private aspect of the user’s life. \\ \hline
        \multirow{1}{1.5cm}{Unawareness} & 13. Private mental health information is accessed by unauthorized parties. \\
        & 14. User is not aware of how and why their mental health data is being collected, processed, stored, and shared. \\ \hline
        \multirow{1}{1.5cm}{Non-compliance} & 15. User is concerned about the processing and storage of mental health data against privacy regulations or policies. \\
        & 16. Mental health data is being exploited for other purposes. \\
        & 17. Mental health data is shared with third parties. \\
        \hline
    \end{tabular}
\end{table}

\textit{Choice of NLI model.}
Several state-of-the-art (SOTA) language models, such as BERT \cite{devlin2019bert}, RoBERTa \cite{liu2019roberta}, and DeBERTa \cite{he2021deberta}, have been used for various Natural Language Understanding (NLU) tasks. DeBERTa outperforms BERT and RoBERTa in most NLU tasks due to its disentangled attention and an enhanced mask decoder \cite{he2021deberta}. The V3 version of DeBERTa further improves its performance significantly by using advanced pre-training approaches \cite{he2021debertav3}. The DeBERTa-V3 base model has 12 layers and a hidden size of 768. Also, it has only 86M backbone parameters with a vocabulary containing 128K tokens, which introduces 98M parameters in the embedding layer. Thus, due to its lightweight and superior performance over BERT and RoBERTa, we decided to use DeBERTa-V3 for our NLI task. We used the variant of this model available in the HuggingFace library \cite{wolf2019huggingface} that is fine-tuned on the three SOTA NLI datasets, namely: Multi-Genre NLI (MNLI) \cite{williams2017broad} (433k sentence pairs), Adversarial NLI (ANLI) \cite{nie2019adversarial} (169k sentence pairs), Fact Extraction and VERification NLI (FeverNLI) \cite{thorne2019fever2} (185k sentence pairs).


\begin{table}[t]
    \renewcommand{\arraystretch}{1.2}
    \caption{Results of the experiments to select the best set of hypotheses in SAGE.}
    \label{tab:r1}
    \begin{tabular}{p{5cm}|p{.6cm}p{.6cm}p{.6cm}}
        \textbf{Heuristics} & \textbf{P} & \textbf{R} & \textbf{F\textsubscript{1}} \\ \hline
        \textbf{Set 1}: \textit{N\textsubscript{E}(i, 0.9)$>$=1} or \textit{N\textsubscript{E}(i, 0.8)$>$=3} or \textit{N\textsubscript{E}(i, 0.75)$>$=5} & 0.34 & 0.70 & 0.46 \\
        \textbf{Set 2}: \textit{N\textsubscript{E}(i, 0.85)$>$=1} or \textit{N\textsubscript{E}(i, 0.75)$>$=3} or \textit{N\textsubscript{E}(i, 0.7)$>$=5} & 0.40 & 0.86 & \textbf{0.55} \\
        \textbf{Set 3}: \textit{N\textsubscript{E}(i, 0.8)$>$=1} or \textit{N\textsubscript{E}(i, 0.7)$>$=3} or \textit{N\textsubscript{E}(i, 0.65)$>$=5} & 0.35 & 0.87 & 0.49 \\
        \textbf{Set 4}: \textit{N\textsubscript{E}(i, 0.75)$>$=1} or \textit{N\textsubscript{E}(i, 0.65)$>$=3} or \textit{N\textsubscript{E}(i, 0.6)$>$=5} & 0.31 & 0.90 & 0.47 \\
        \hline
    \end{tabular}
    \medskip
    
    \noindent Set 2 heuristics selected in SAGE:
    \begin{itemize}
        \item A review i is labeled as \textit{maybe-privacy} if \textit{N\textsubscript{E}(i, 0.85)$>$=1} or \textit{N\textsubscript{E}(i, 0.75)$>$=3} or \textit{N\textsubscript{E}(i, 0.7)$>$=5}; where \textit{N\textsubscript{E}(i, t)} is the number of hypotheses receiving an \textit{entailment\_score} above a threshold \textit{t} for review \textit{i}.
    	\item The rest of the reviews are labeled as \textit{maybe-not-privacy}.
    \end{itemize}
\end{table}

\textit{Defining heuristics.}
Heuristics are the thresholds that are used to select potential privacy reviews based on the \textit{entailment\_scores} of the corresponding privacy hypotheses. To determine the best set of heuristics, taking inspiration from \cite{harkous2022hark, duvsek2020evaluating}, we followed the intuition-based method, where an app review is potentially considered within the privacy domain if it satisfies a greater number of hypotheses with a high \textit{entailment\_score}; the higher the score, the more probable it is that the review is privacy related. Thus, using this intuition method, we performed experiments with 4 different sets of heuristics (shown in Table \ref{tab:r1}) ranging from high to low values of \textit{entailment\_score} and selected the best set of heuristics that yielded the minimum number of false positives (FP) (0-labeled reviews annotated as `maybe-privacy') and false negatives (FN) (1-labeled reviews annotated as `maybe-not-privacy'). We used the F1 score to perform this analysis.

Table \ref{tab:r1} shows the experiment results and highlights that Set 2 achieved the highest F1 score of 0.54. However, the F1 score of all sets was within the range of 0.45-0.55 due to the NLI model's propensity to generate a large number of FPs, thereby exhibiting low precision while still assisting in filtering out irrelevant data. We also observed that a decrease in threshold values resulted in a minor increase in true positives (TP), but there was a significant rise in FPs (further details are available in our replication package). Consequently, the Set 2 heuristics were chosen for subsequent phases in SAGE.


\textbf{\underline{B) LLM Inference}} 
We leveraged the GPT model to extract privacy reviews from a set of potential privacy reviews identified during the NLI inference. First, we designed the prompt and configured the LLM settings (step \circled{E}). Then, we used the GPT model and performed the inference operation (step \circled{F}) on a set of \textit{`maybe-privacy'} reviews (potential privacy reviews) extracted in the NLI step. We ensured to prompt the LLM five times before selecting the majority response for each review; as a result, we obtained the \textit{yes} (privacy) and \textit{no} (not-privacy) labeled dataset of app reviews. 

\begin{figure}[b]
    \centering
    \includegraphics[scale=0.16]{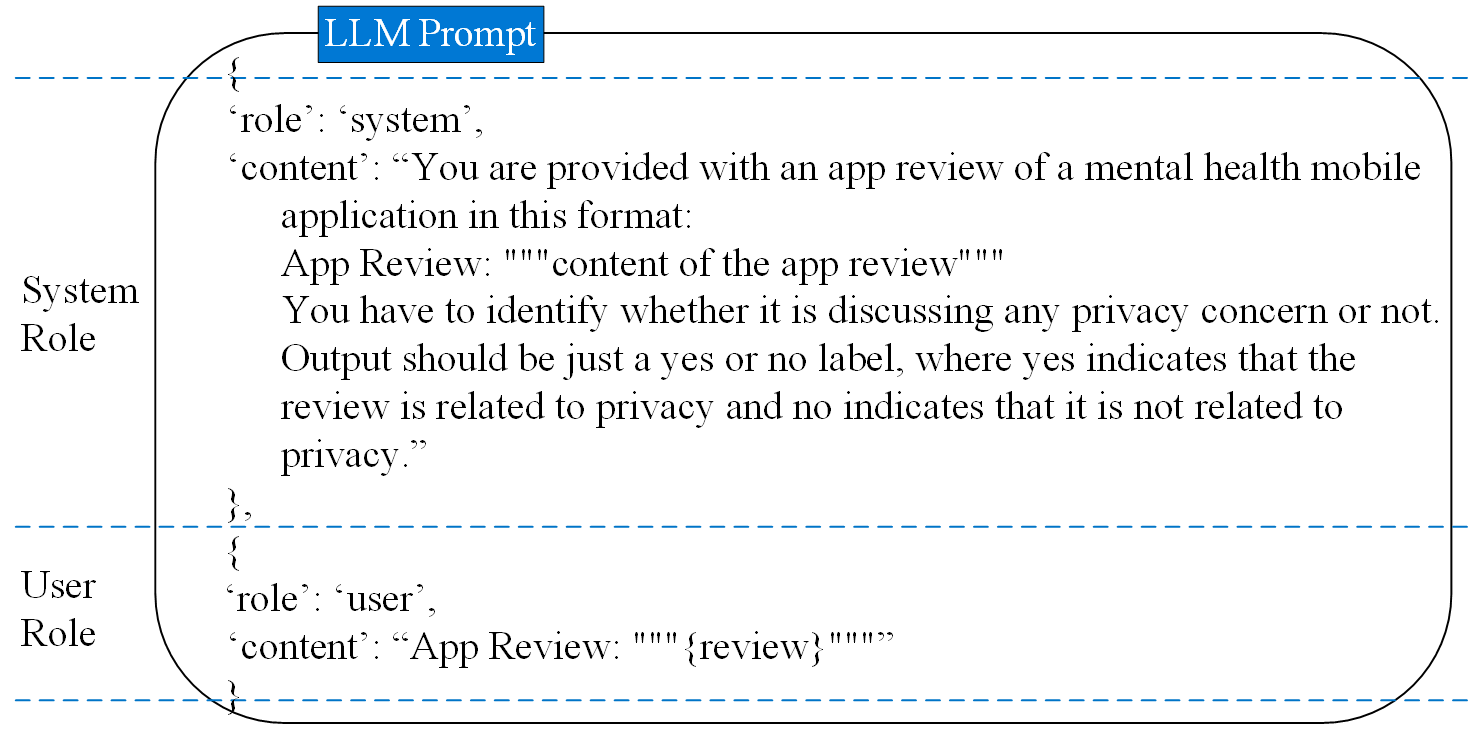}
    \caption{Role-based prompt designed by following the guidelines from \cite{chen2023unleashing}. The system role is used to give instructions to the LLM, and the user role is to provide the app review and get the response.} 
    \label{fig:prompt}
\end{figure} 

\textit{Prompt design and LLM configuration.} In line with previous works \cite{touvron2023llama, zhang2023revisiting}, we designed a prompt for the GPT model to experiment with the zero-shot setting and followed the guidelines from \cite{chen2023unleashing} to design the prompt and configure the temperature of the LLM. The temperature plays a crucial role in generating responses as it controls the randomness of the generated output, where a lower temperature leads to more deterministic outputs \cite{chen2023unleashing}. Thus, we set the temperature parameter to 0 to minimize the likelihood of the model generating incorrect or unpredictable outputs. Additionally, we prompted the model five times and selected the majority response to overcome the inherent variability in the model’s responses and increase the chances of obtaining an improved deterministic output \cite{chen2023unleashing}.

To design the prompt, we followed the role prompting technique by defining clear and precise instructions for each role. This technique involves giving the model a specific role, such as a helpful assistant or an expert \cite{chen2023unleashing}. It can be particularly effective in guiding the responses of the model and ensuring that they align with the desired output \cite{chen2023unleashing}. In our design, we defined two roles, system, and user, for the model. In the system role, we first defined the format of the input provided to the model, comprising of an app review as \textit{App Review: ``````content of the app review"""}. Next, we described the classification task to instruct the model on what to do, and at the last, we provided the desired format for the model output, where we specifically asked the model to return just the \textit{yes}/\textit{no} labels indicating whether the input app review is related to privacy or not. In the user role, we provided the model with the app review in the desired format. Figure \ref{fig:prompt} shows our detailed prompt structure. 

\textit{Choice of GPT LLM.} Several GPT LLMs have recently been proposed, with a prevalence of studies using the GPT3.5-turbo \cite{ouyang2023llm}. However, GPT-4o-mini (the model that powers ChatGPT at the time of writing this) is relatively affordable, capable, and fast, and is an improved version of GPT3.5-turbo \cite{ye2023comprehensive}. Therefore, we decided to use GPT-4o-mini \cite{hurst2024gpt} to evaluate the efficacy of LLMs in classifying privacy reviews. GPT-4o-mini has the knowledge cutoff of October 2023, a context window of 128K tokens, and 16,383 max output tokens. We used the OpenAI API for all our experiments, and there was only a single version of GPT-4o-mini available on the OpenAI API i.e. \textit{gpt-4o-mini-2024-07-18}.

\begin{figure}[t]
    \centering
    \includegraphics[scale=0.27]{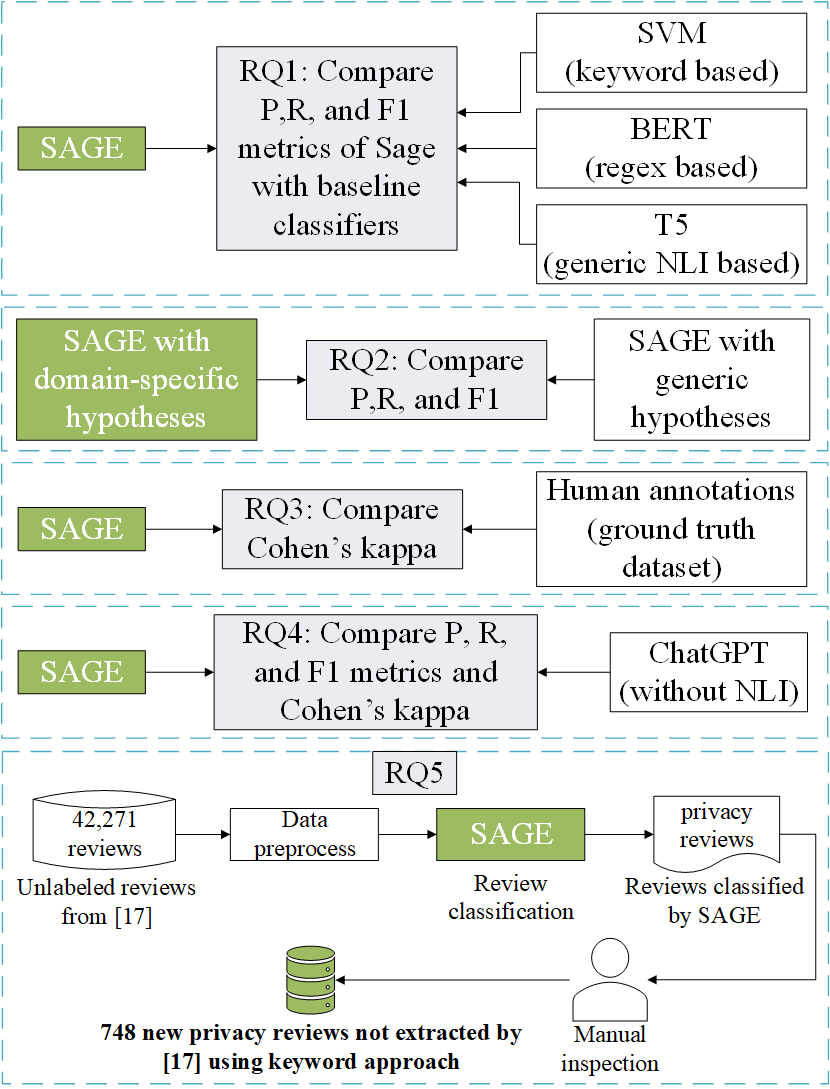}
    \caption{SAGE evaluation process. RQ1-RQ4 represent the quantitative evaluation of SAGE, and RQ5 represents the qualitative evaluation of SAGE.}
    \label{fig:evaluation}
\end{figure}

\section{Evaluation design} \label{evaluation}
\textbf{\underline{A) Research questions.}} 
To assess the validity of SAGE, our evaluation focuses on the following RQs:

\textit{RQ1: What is the effectiveness of SAGE compared to the baseline privacy review classifiers?} 
Previous studies have used SVM \cite{nguyen2019short}, BERT \cite{nema2022analyzing}, and T5 \cite{harkous2022hark} models for the classification of privacy reviews. These models vary across architecture dimensions and require fine-tuning on the labeled datasets. Additionally, these studies have used keyword \cite{nguyen2019short}, regex \cite{nema2022analyzing}, or generic NLI \cite{harkous2022hark} approaches for training data sampling. For these reasons, we selected these models as our baseline classifiers.

\textit{RQ2: What is the effectiveness of SAGE compared to generic privacy hypotheses?} We investigate the effectiveness of our MH domain-specific privacy hypotheses compared to generic privacy hypotheses utilized by Harkous et al. \cite{harkous2022hark}. This analysis will allow us to gain deeper insight into how domain-specific hypotheses yield fewer FP due to the domain dependence of users' privacy concerns.

\textit{RQ3: To what extent can we leverage SAGE to reduce manual data labeling efforts?} Inspired by previous work on prompt-based data labeling in low-resource settings \cite{wang2021want}, we investigate to what extent we can leverage SAGE to reduce the efforts associated with human annotation to build reliable gold-standard datasets.

\textit{RQ4: What is the effectiveness of SAGE compared to ChatGPT (without NLI)?} Our goal is to assess the efficacy of NLI in SAGE by comparing it with ChatGPT, which does not include the NLI step and does not require fine-tuning, similar to SAGE. This will help in understanding to what extent NLI is successful in filtering irrelevant app reviews in SAGE.

\textit{RQ5: What is the effectiveness of SAGE compared to the traditional keyword-based method?} We perform a qualitative evaluation of SAGE and try to extract privacy reviews that do not contain any predefined keywords.

\textbf{\underline{B) Evaluation dataset}}
We utilized the ground truth data (manually validated) from Ebrahimi et al. \cite{ebrahimi2022unsupervised}, consisting of 1,376 (414 privacy and 962 not-privacy reviews) manually annotated reviews. Table \ref{tab:stat} shows statistical information about the dataset. This dataset was developed using keyword-based filtering alongside manual inspection of over 204K reviews mined from the most widely used MH apps available on the Google Play Store and Apple App Store. The data collection and labeling process are presented in detail in the study of Ebrahimi et al. \cite{ebrahimi2022unsupervised}. For the quantitative evaluation (RQ1-RQ4) of SAGE, we used the labeled dataset of 1,376 reviews. For the qualitative evaluation (RQ5), we used the unlabeled dataset of 42,271 reviews and extracted new privacy reviews.



\begin{table}[t]
    \centering
    \renewcommand{\arraystretch}{1.1}
    \caption{Statistics of the dataset used from \cite{ebrahimi2022unsupervised}.}
    \label{tab:stat}
    \vspace{-2mm}
    \begin{tabular}{l|p{3.5cm}}
        \textbf{Number of apps} & 5 (Calm, Headspace, Sanvello, Talkspace, and Shine) \\
        \hline
        \textbf{App domain} & Health \& Fitness (MH) \\
        \hline
        \textbf{Total reviews} & 204,374 \\
        \hline
        \textbf{1-2 star rated reviews} & 43,647 \\
        \hline
        \textbf{Privacy labeled reviews} & 414 \\
        \hline
        \textbf{Not-privacy labeled reviews} & 962 \\
        \hline
        \textbf{Unlabeled reviews} & 42,271 \\
        \hline
        \textbf{Average number of words per review} & 33 \\
        \hline
        \textbf{Time range} & 2012-01-07 to 2021-10-06
    \end{tabular}
\end{table}

\textbf{\underline{C) Evaluation methods}}
Figure \ref{fig:evaluation} shows the evaluation design for SAGE, structured according to our RQs. To address RQ1, RQ2, and RQ4, we assessed SAGE in terms of Precision (P), Recall (R), and F1. This choice reflects the standard methodology to assess the performance of text categorization approaches \cite{sebastiani2002machine} and is consistent with previous works \cite{nema2022analyzing}. P is the ratio between the true positives (TP) and all the predicted items for a given class. R represents the ratio of TP and all items belonging to a given polarity class. The F1 is computed as the harmonic mean of precision and recall. We used the macro-averaged values of P, R, and F1, thereby enabling a quick comparison of the overall performance of SAGE with the baselines. Furthermore, we computed Cohen's $kappa$ \cite{cohen1968weighted} to assess the agreement between SAGE and human annotators to address RQ3 and RQ4. To address RQ5, we performed the qualitative analysis on the unlabeled dataset and extracted new privacy reviews to show the efficacy of SAGE on unseen datasets. We provide details of the baselines for each RQ and discuss our evaluation plan in detail as follows. 

\noindent
\textbf{Privacy classifiers (RQ1):}
We compared SAGE with three baseline supervised privacy classifiers.

\textit{Support Vector Machines (SVM)}. A traditional machine learning model, effective for text classification \cite{joachims1998text}, especially for short documents \cite{basu2003support}. We used the SVM model based on the bag of words approach (using 3-5 character n-grams), reproducing the one used in Nguyen et al. \cite{nguyen2019short}. 

\textit{Bidirectional Encoder Representations from Transformers (BERT)}. An encoder-only language representation model based on the transformer architecture, trained on a 3.3 billion word corpus \cite{devlin2019bert}. We fine-tuned the pre-trained BERT model on our dataset for three epochs and chose the best epoch model based on the performance of the validation dataset, reproducing the one used in Nema et al. \cite{nema2022analyzing}.

\textit{Text-to-Text Transfer Transformer (T5)}. An encoder-decoder model, with a unified architecture based on casting problems into the text-to-text paradigm and training a model on the text generation objective. We cast the task of classifying app reviews into text-to-text generation and fine-tuned the T5 model, reproducing the one used in Harkous et al. \cite{harkous2022hark}.

We used the 80-20 dataset split (meaning 80\% data for training and 20\% data for testing or validating) with stratified distribution over both classes. We reported P, R, and F1 metrics of all the models and compared them with SAGE.

\noindent
\textbf{Generic Hypotheses (RQ2):} We assessed the validity of our MH domain-specific privacy hypotheses by comparing them with the set of 31 generic hypotheses (not related to any particular domain) and corresponding heuristics (both available in our replication package) provided by Harkous et al. \cite{harkous2022hark}. These hypotheses were derived from Solove's \cite{solove2005taxonomy} and Wang and Kobsa's \cite{wang2009privacy} taxonomies for privacy violations and privacy-enhancing technologies. For evaluation, we replaced the domain-specific hypotheses in SAGE with the generic hypotheses, repeated all the experiments, and compared the performance using P, R, and F1 metrics.

\noindent
\textbf{Human Annotation (RQ3):} We assessed the validity of SAGE in aiding the creation of reliable gold-standard datasets. We reported Cohen's $kappa$ measure \cite{cohen1968weighted}, which represents the agreement level between the annotation achieved by SAGE and the true labels annotated by humans.

\noindent
\textbf{ChatGPT (RQ4):} We assessed the efficiency of NLI in SAGE by comparing it with ChatGPT, which did not include the NLI step. We used the same settings and prompt for ChatGPT; only the data changed, as we did not perform the NLI inference, so we used all 1,376 labeled reviews for this experiment. We reported P, R, and F1 metrics to compare the performance for the classification task and Cohen's $kappa$ to compare the agreement level with human annotations.

\noindent\textbf{Keyword-based method (RQ5):} We performed the qualitative evaluation of SAGE by comparing it with the traditional keyword-based method \cite{ebrahimi2022unsupervised}. Here, we tried to achieve one of the important objectives of SAGE, i.e., extracting privacy reviews that do not contain predefined keywords. We utilized the unlabeled dataset of 42,271 app reviews and first preprocessed all the reviews to remove special characters, emoticons, and white spaces, and converted them to lowercase. Next, we extracted privacy reviews using SAGE. Then, we used manual inspection (detailed below) to create a ground truth dataset of previously unidentified privacy reviews.

\textit{Manual inspection setup:}
We performed a manual inspection to scrutinize each extracted review for privacy concerns. Four annotators, including the first author and three graduate students from our research lab, conducted this task. The first author analyzed all the reviews, while the others simultaneously inspected one-third of the sample, such that each review was inspected at least twice. To prevent exhaustion, we performed this process in a 7-day timeframe. To ensure the understanding of the task and the definitions for privacy and not-privacy labels, we created the labeling instructions (available in our replication package), and we based our analysis on the privacy concepts provided by Iwaya et al. \cite{iwaya2023privacy} to label the reviews. After the manual inspection, we cross-checked the findings of the manual classification. For every disagreement, a third annotator was requested to break the tie. In total, 137 reviews had to be further analyzed by another annotator. To determine the extent to which the annotators agreed upon the classifications, we used Cohen’s $kappa$ coefficient \cite{cohen1968weighted}. We acquired a degree of agreement of 0.74, which shows substantial agreement and indicates high reliability in the annotation process \cite{viera2005understanding}.

\textbf{\underline{D) Computational Resources}}
The experiments were conducted on an NVIDIA GeForce RTX 4090 GPU of 40 GB of RAM and a 24-core CPU setup. We implemented our models using Python 3.12 with CUDA version 11.8 and HuggingFace Transformers version 4.44.1. We used NumPy and Pandas for linear algebra operations. 

\section{Evaluation results} \label{results}
In this section, we explain the results of our RQs.

\begin{table}[t]
    \centering
    \caption{\textbf{Results of quantitative evaluation of SAGE (RQ1-RQ4).}}
    \label{tab:results}
    \begin{tabular}{p{4cm}|p{.6cm}p{.6cm}p{.6cm}p{1cm}}
        \textbf{Model} & \textbf{P} & \textbf{R} & \textbf{F1} & \textbf{Cohen's $kappa$} \\ \hline
        \rowcolor{green!15} SAGE (our approach) & 0.84 & 0.86 & 0.85 & 0.71 \\
        SVM (RQ1) & 0.92 & 0.91 & 0.91 & - \\
        BERT (RQ1) & 0.86 & 0.83 & 0.84 & - \\
        T5 (RQ1) & 0.83 & 0.80 & 0.79 & - \\
        SAGE with generic hypotheses (RQ2) & 0.75 & 0.79 & 0.77 & - \\
        ChatGPT (RQ4) & 0.72 & 0.80 & 0.76 & 0.62 \\
        \hline
    \end{tabular}
\end{table}

\textbf{Answering RQ1:} Table \ref{tab:results} presents the comparison of SAGE with baseline privacy classifiers: SVM (Nguyen et al. \cite{nguyen2019short}), BERT (Nema et al. \cite{nema2022analyzing}), and T5 (Harkous et al. \cite{harkous2022hark}). We can observe that SVM is the best-performing classifier in this setting, with an F1 score of 0.91, whereas T5 receives the lowest F1 score of 0.79. SAGE and BERT have shown almost the same performance, with an F1 score of 0.85 and 0.84, respectively. Our dataset is comparatively small and contains privacy-indicative words, due to which SVM can learn the patterns well and has shown better performance. However, we highlight that SAGE is not fine-tuned on the labeled dataset like other baseline classifiers and has still achieved notable results. We also note that using the base model of T5 as compared to the much larger variant T5-11B (used in the baseline study \cite{harkous2022hark}) results in lower performance.
\begin{mdframed}
SAGE accurately classified privacy reviews with zero-shot learning (without fine-tuning on the labeled dataset), achieving an F1 score of 0.85, outperforming BERT and T5 baselines.
\end{mdframed}

\textbf{Answering RQ2. } For the RQ2 evaluation, we replaced the domain-specific hypotheses in SAGE with generic hypotheses. From Table \ref{tab:results}, it can be observed that SAGE with domain-specific hypotheses yields better performance than generic hypotheses. Additionally, at the NLI step of SAGE, we observed that generic hypotheses yield more FP (total of 1,150 `maybe-privacy' reviews) compared to domain-specific hypotheses (total of 926 `maybe-privacy' reviews). This means that generic hypotheses extracted more irrelevant information, which was further used in LLM inference, yielding poor performance compared to the domain-specific hypotheses. Detailed results are included in our replication package.
\begin{mdframed}
SAGE outperformed generic hypotheses, highlighting the importance of using MH domain-specific privacy hypotheses in our approach to provide domain-specific contextual awareness while classifying privacy reviews.
\end{mdframed}

\textbf{Answering RQ3: }Table \ref{tab:results} shows the agreement level between SAGE and human annotators using Cohen's $kappa$ to assess the labeling ability of SAGE. For the interpretation of $kappa$ values, we follow a consolidated interpretation \cite{viera2005understanding} suggesting that the agreement is less than chance if $k \leq 0$, slight if $0.01 \leq k \leq 0.20$, fair if $0.21 \leq k \leq 0.40$, moderate if $0.41 \leq k \leq 0.60$, substantial if $0.61 \leq k \leq 0.80$, and almost perfect if $0.81 \leq k \leq 1$. According to this interpretation, we observed a substantial agreement, with a value of $k = 0.71$, between SAGE and human annotations (ground truth dataset).
\begin{mdframed}
SAGE showed a substantial level of agreement with human annotations, achieving Cohen's $kappa$ of 0.71.
\end{mdframed}

\textbf{Answering RQ4:} From Table \ref{tab:results}, it can be observed that ChatGPT shows poor performance compared to SAGE, with an F1 score of 0.72. This lower performance is due to the large volume of irrelevant information that is fed to the LLM without filtering the dataset using NLI. Additionally, we report the agreement level between ChatGPT and human annotations using Cohen's $kappa$, as shown in Table \ref{tab:results}. We can observe that the agreement is still substantial, but the level has decreased compared to the agreement level between SAGE and human annotators.
\begin{mdframed}
SAGE showed superior performance compared to ChatGPT, highlighting the importance of using NLI to filter irrelevant information.
\end{mdframed}

\textbf{Answering RQ5:} To extract a new set of privacy reviews from the dataset of 42,271 unlabeled reviews, we first preprocessed all the reviews and then performed the classification task using SAGE. At the intermediary stage of SAGE, i.e., after NLI inference, we identified 4,591 `maybe-privacy’ reviews. These reviews were then used in LLM inference operation, and 1,155 reviews were further labeled as \textit{yes} by the GPT model, indicating the privacy reviews. After this, we performed the manual inspection and found 748 new privacy reviews (new ground truth dataset) that were not extracted by Ebrahimi et al. \cite{ebrahimi2022unsupervised} using keyword-based filtering. We show a few examples of new privacy reviews in Section \ref{discuss}.
\begin{mdframed}
SAGE extracted 748 privacy reviews without any fine-tuning from a dataset of more than 42K app reviews.
\end{mdframed}

\section{Discussion} \label{discuss}


\textbf{SAGE vs. Supervised classifiers and ChatGPT.} We compared the performance of SAGE with SVM \cite{nguyen2019short}, BERT \cite{nema2022analyzing}, and T5 \cite{harkous2022hark}. While not directly comparable due to the difference in datasets in the respective studies, we trained these models on the ground truth dataset used in our study for the comparison with SAGE. SAGE achieved an F1 score of 0.85, closely trailing BERT's 0.84 and surpassing T5's 0.78. Notably, while all baseline models underwent fine-tuning for this task, SAGE operated in a zero-shot setting. This implies SAGE's capability to classify privacy reviews effectively without specific fine-tuning, presenting a clear advantage when labeled data is scarce. Furthermore, we emphasize the role of NLI in eliminating irrelevant information, illustrated by the comparison of SAGE with ChatGPT. App reviews contain extensive irrelevant content, with only 0.5\% pertaining to privacy \cite{mukherjee2020empirical}, making it impractical to input all irrelevant data into our GPT classifier, as evidenced by our results (Table \ref{tab:results}).

\textbf{Reducing the manual efforts associated with data labeling.} The development and maintenance of software modules utilizing LLMs present challenges despite their significant potential. Deploying LLMs for classification may be impractical when computational resources are scarce. In such scenarios, the use of traditional machine learning classifiers, such as SVM, may prove advantageous. However, this necessitates the creation of a gold-standard dataset for training purposes. Our study reveals a Cohen’s kappa of $0.71$, indicating considerable agreement between SAGE and human annotators. Then, in line with previous work \cite{zhang2023revisiting, zhu2023can}, we recommend integrating SAGE into an annotation team to mitigate the manual efforts of human annotation \cite{wang2021want} or to assist with data augmentation \cite{moller2023parrot}. Alternatively, it may serve to sample data for human annotation from a large corpus of app reviews, which often includes considerable irrelevant content (shown in RQ5).

\begin{table*}[h]
    \renewcommand{\arraystretch}{1.2}
    \centering
    \caption{Examples of privacy reviews extracted using SAGE. None of the reviews contains privacy-indicative keywords.}
    \label{tab:rr}
    \begin{tabular}{llp{12.5cm}}
         \textbf{App} & \textbf{Date (DD/MM/YY)} & \textbf{Privacy Review}  \\
         \hline
         Talkspace & 06/06/2016 & She read all the notes from my previous therapy! This is unethical.\\
         Headspace & 10/04/2021 & Therapy notes being accessed and shared with the world.\\
         Shine & 05/12/2019 & How are you different from any other app now that is interested in our user patterns over our mental health?\\
         Sanvello & 07/05/2017 & New update deleted all my journals and mood tracking, and not able to recover.\\
         Calm & 09/07/2021 & The app collects data about websites you visit "before or after the use of our services".\\
         \hline
    \end{tabular}
\end{table*}

\textbf{New ground truth dataset created using SAGE.} Using SAGE, we extracted 748 new privacy reviews that did not contain predefined keywords. 
We present a few reviews in Table \ref{tab:rr} and make the whole dataset publicly available. All these reviews discuss users' privacy concerns and are specifically related to the MH domain, including concerns related to personal mental health data, such as therapy notes and private chats with therapists. 
Further, we executed an automated analysis utilizing wordcloud generation and bigram extraction methodologies, which revealed significant privacy concerns associated with sensitive information related to credit card and banking details, particularly in the context of the subscription model employed by various apps. 
This preliminary analysis shows the potential of SAGE in extracting privacy reviews, which can be refined into actionable requirement artifacts as shown by Sorathiya et al. \cite{sorathiya2024towards} to address users' privacy concerns. Our replication package includes a detailed analysis of these concerns.


\textbf{Abstract tool for app developers and adaptation of SAGE to other domains.} SAGE is an abstract tool comprising two primary components: NLI with MH domain-specific hypotheses and a GPT model. This tool helps app developers analyze user feedback and address their concerns. Furthermore, we highlight that SAGE is adaptable to other domains, such as the sharing economy, finance, or gaming, to extract reviews related to ethical concerns like safety, transparency, or accountability. Additionally, it can be adapted to analyze feedback from social media platforms such as Reddit \cite{li2022narratives}. This adaptability is achieved by defining hypotheses related to any ethical concern within any app domain and incorporating them into SAGE, showcasing its reusability and adaptability across domains.

\section{Limitations and Threats to Validity} \label{threats}
\textbf{Construct threats}: 
Dataset development is labor-intensive and susceptible to reader bias. To mitigate this, we implemented a systematic manual inspection methodology involving four annotators to minimize individual bias. Additionally, we employed SOTA NLI and GPT models in SAGE; however, we encourage future studies to experiment with a different set of models. The evaluation metrics P, R, and F1 are well-established in the SE domain and recommended for such tasks.

\textbf{Internal threats}: 
The domain-specific privacy hypotheses and heuristics, together with GPT prompt design and parameter settings, may pose internal validity threats. We used established techniques to mitigate these threats. Consistent with previous work \cite{harkous2022hark, fazelnia2024lessons}, we developed privacy hypotheses based on prevalent MH domain privacy concepts \cite{iwaya2023privacy}. Additionally, we followed the methodology of past research \cite{harkous2022hark, duvsek2020evaluating} to define our heuristics and tested four sets to minimize the bias; however, we recognize that altering these could affect results. The GPT model prompt was developed per the standards of \cite{chen2023unleashing}. Furthermore, a low model temperature was set for deterministic outputs from the GPT model, as high temperatures could lead to unpredictable results. Future research may explore the effects of varying this parameter in classification tasks.

\textbf{External threats}:
The focus on a GPT model from OpenAI constrains the generalizability of our results. We advocate in favor of future replications, also exploring open-source models such as LLaMa3.1 \cite{dubey2024llama}, Falcon \cite{almazrouei2023falcon}, and Mistral \cite{jiang2023mistral}. Future studies can explore few-shot learning to reduce the risk of model bias and hallucination. Future replications with larger and more varied benchmark datasets, including app reviews from different domains and platforms, are also required. These will enable evaluating the generalizability of our findings, thus assessing how consistent the behavior of SAGE is across different models and datasets.

\section{Related Work} \label{brw}


\noindent \textbf{Privacy in app reviews. }
The prior studies closest to ours are presented in Table \ref{tab:rw}. The majority of existing work employs traditional NLP approaches to develop classifiers and rely on keyword or regex-based approaches to perform training data sampling for these classifiers \cite{besmer2020investigating, mukherjee2020empirical, nguyen2019short, ebrahimi2022unsupervised, nema2022analyzing}. These choices are intercorrelated, as it is easy to achieve high performance using traditional models on test sets created with such sampling methods \cite{harkous2022hark}. However, their reliance on pre-defined, context-independent keywords may not reflect the actual terminology used by reviewers and potentially lead to inaccuracies \cite{alomar2021finding}.
Harkous et al. \cite{harkous2022hark} utilized the NLI method to alleviate the limitations of keyword-based search techniques, conducting a detailed analysis of users’ privacy concerns in app reviews through established privacy taxonomies \cite{solove2005taxonomy, wang2009privacy}. They developed 31 privacy hypotheses for the NLI task, yet their approach predominantly employed generic privacy concepts, which can fail to identify MH domain-specific privacy concerns as users' privacy concerns are domain-dependent \cite{ebrahimi2022unsupervised}. Moreover, all these studies rely on labeled data for fine-tuning classification models, highlighting a limitation in scenarios with limited or no labeled data.

\begin{table}[t]
    \centering
    \renewcommand{\arraystretch}{1.2}
    \caption{Details of related works in comparison to SAGE.}
    \label{tab:rw}
    \begin{tabular}{p{2cm}p{0.8cm}p{1.2cm}p{1.2cm}p{1.5cm}}
        \textbf{Study} & \textbf{Domain} & \textbf{Data Sampling} & \textbf{Classifier Model} & \textbf{Fine-tuning Required?} \\
        \hline
        Besmer et al. \cite{besmer2020investigating} & General & Keywords & LR & Yes \\
        Mukherjee et al. \cite{mukherjee2020empirical} & General & Keywords & SVM & Yes \\
        Nguyen et al. \cite{nguyen2019short} & General & Keywords & SVM & Yes \\
        Ebrahimi et al. \cite{ebrahimi2022unsupervised} & MH & Keywords & - & - \\
        Nema et al. \cite{nema2022analyzing} & General & Regex & BERT & Yes \\
        Harkous et al. \cite{harkous2022hark} & General & Generic NLI & T5 & Yes \\
        \rowcolor{green!15} SAGE (our approach) & MH & Domain-specific NLI & GPT & No \\
        \hline
    \end{tabular}
\end{table}

\textbf{Our methodology} differs from previous studies in multiple aspects. Primarily, we focus on privacy within the MH app domain. Secondly, we employ NLI for sampling privacy reviews, contrasting with the keyword or regex methods adopted by others \cite{besmer2020investigating, mukherjee2020empirical, nguyen2019short, ebrahimi2022unsupervised, nema2022analyzing}. Moreover, we apply MH domain-specific privacy hypotheses for NLI rather than the generic ones previously used in \cite{harkous2022hark}. Third, we utilize the GPT model to classify privacy reviews, as opposed to conventional supervised classification techniques \cite{nguyen2019short, nema2022analyzing, harkous2022hark}. Additionally, our method eliminates the need for fine-tuning on labeled datasets, thereby minimizing manual data labeling efforts. Ultimately, we emphasize that our objective is to extract privacy reviews to identify privacy requirements, which distinguishes our focus from other studies that do not engage in RE-related tasks.


\noindent \textbf{LLMs}: 
LLMs are categorized into three groups based on their architecture structure: 1) encoder-only LLMs (Eg: BERT), 2) encoder-decoder LLMs (Eg: T5), and 3) decoder-only LLMs (Eg: GPT) \cite{hou2023large}. Recently, LLMs have gained prominence in software engineering (SE) due to their problem-solving capabilities across various applications \cite{hou2023large}. 
More recently, from 2023, decoder-only LLMs have emerged prominently in SE research and have been utilized for multiple SE tasks \cite{colavito2024leveraging}. 
In the RE domain, decoder-only LLMs, especially GPT, have been widely used for various tasks, such as requirements classification \cite{arora2024advancing}, generation of new requirements \cite{bencheikh2023exploring}, requirements elicitation \cite{ronanki2023investigating, white2024chatgpt}, requirements analysis \cite{rasheed2024autonomous}, and resolving ambiguity in requirements \cite{sridhara2023chatgpt}. ChatGPT has shown benefits with enhanced productivity and cost savings for various RE tasks \cite{marques2024using}. Despite such promising results for various RE tasks, GPT LLM has not yet been leveraged in the domain of RE for privacy review classification, which we address in our study.

\section{Conclusion} \label{conclude}
Mining privacy reviews is a key to eliciting privacy requirements and addressing users' privacy concerns. To date, a domain-specific approach that can effectively identify privacy reviews without heavily relying on labeled data or predefined keywords has been lacking. This paper is the first to provide a framework (SAGE) that automates the extraction of privacy reviews using domain-specific context awareness (using NLI) and without any fine-tuning on the labeled dataset (using GPT). Beyond its immediate application in privacy requirements engineering, SAGE could be adapted to other domains and ethical concerns, such as safety and transparency, with modifications to hypotheses. We believe that SAGE advances the state-of-the-art towards a critical step in the requirements elicitation process by leveraging user feedback in app reviews.

\section{Acknowledgement}
This research is supported and funded by the NSERC Alliance-Alberta Innovates Advance Program Stream I program. Also, we thank the anonymous reviewers for their constructive feedback, which has helped improve this work further.

\bibliographystyle{IEEEtran}
\bibliography{espre_ref}

\begin{thebibliography}{10}
\providecommand{\url}[1]{#1}
\csname url@samestyle\endcsname
\providecommand{\newblock}{\relax}
\providecommand{\bibinfo}[2]{#2}
\providecommand{\BIBentrySTDinterwordspacing}{\spaceskip=0pt\relax}
\providecommand{\BIBentryALTinterwordstretchfactor}{4}
\providecommand{\BIBentryALTinterwordspacing}{\spaceskip=\fontdimen2\font plus
\BIBentryALTinterwordstretchfactor\fontdimen3\font minus \fontdimen4\font\relax}
\providecommand{\BIBforeignlanguage}[2]{{%
\expandafter\ifx\csname l@#1\endcsname\relax
\typeout{** WARNING: IEEEtran.bst: No hyphenation pattern has been}%
\typeout{** loaded for the language `#1'. Using the pattern for}%
\typeout{** the default language instead.}%
\else
\language=\csname l@#1\endcsname
\fi
#2}}
\providecommand{\BIBdecl}{\relax}
\BIBdecl

\bibitem{reinert2024state}
M.~Reinert \emph{et~al.}, ``The state of mental health in america 2024,'' 2024.

\bibitem{bauer2020smartphones}
M.~Bauer \emph{et~al.}, ``Smartphones in mental health: a critical review of background issues, current status and future concerns,'' \emph{International journal of bipolar disorders}, vol.~8, pp. 1--19, 2020.

\bibitem{eagle2022money}
T.~Eagle \emph{et~al.}, ``" money doesn't buy you happiness": negative consequences of using the freemium model for mental health apps,'' \emph{Proc. of the ACM on Human-Computer Interaction}, vol.~6, no. CSCW2, pp. 1--38, 2022.

\bibitem{torous2018dichotomies}
J.~Torous \emph{et~al.}, ``Dichotomies in the development and implementation of digital mental health tools,'' \emph{Psychiatric Services}, vol.~69, no.~12, pp. 1204--1206, 2018.

\bibitem{mcniel2018current}
D.~E. McNiel \emph{et~al.}, ``Current regulation of mobile mental health applications,'' \emph{J Am Acad Psychiatry Law}, vol.~46, pp. 204--11, 2018.

\bibitem{torous2017needed}
J.~Torous \emph{et~al.}, ``Needed innovation in digital health and smartphone applications for mental health: transparency and trust,'' \emph{JAMA psychiatry}, vol.~74, no.~5, pp. 437--438, 2017.

\bibitem{khalid2014mobile}
H.~Khalid \emph{et~al.}, ``What do mobile app users complain about?'' \emph{IEEE software}, vol.~32, no.~3, pp. 70--77, 2014.

\bibitem{aamir2023government}
T.~Aamir, M.~B. Chhetri, M.~Chamikara, and M.~Grobler, ``Government mobile apps: Analysing citizen feedback via app reviews,'' in \emph{2023 38th IEEE/ACM International Conference on Automated Software Engineering (ASE)}.\hskip 1em plus 0.5em minus 0.4em\relax IEEE, 2023, pp. 1858--1863.

\bibitem{ebrahimi2021mobile}
F.~Ebrahimi \emph{et~al.}, ``Mobile app privacy in software engineering research: A systematic mapping study,'' \emph{Information and Software Technology}, vol. 133, p. 106466, 2021.

\bibitem{dkabrowski2023mining}
J.~{Dabrowski}, E.~Letier, A.~Perini, and A.~Susi, ``Mining and searching app reviews for requirements engineering: Evaluation and replication studies,'' \emph{Information Systems}, vol. 114, p. 102181, 2023.

\bibitem{dkabrowski2022analysing}
J.~Dabrowski \emph{et~al.}, ``Analysing app reviews for software engineering: a systematic literature review,'' \emph{Empirical Software Engineering}, vol.~27, no.~2, p.~43, 2022.

\bibitem{sorathiya2024towards}
A.~Sorathiya and G.~Ginde, ``Towards extracting ethical concerns-related software requirements from app reviews,'' in \emph{Proceedings of the 39th IEEE/ACM International Conference on Automated Software Engineering}, 2024, pp. 2251--2255.

\bibitem{mcilroy2016analyzing}
S.~McIlroy, N.~Ali, H.~Khalid, and A.~E.~Hassan, ``Analyzing and automatically labelling the types of user issues that are raised in mobile app reviews,'' \emph{Empirical Software Engineering}, vol.~21, pp. 1067--1106, 2016.

\bibitem{besmer2020investigating}
A.~R. Besmer, J.~Watson, and M.~S. Banks, ``Investigating user perceptions of mobile app privacy: An analysis of user-submitted app reviews,'' \emph{International Journal of Information Security and Privacy (IJISP)}, vol.~14, no.~4, pp. 74--91, 2020.

\bibitem{nema2022analyzing}
P.~Nema, P.~Anthonysamy, N.~Taft, and S.~T. Peddinti, ``Analyzing user perspectives on mobile app privacy at scale,'' in \emph{Proc. of the 44th International Conference on Software Engineering}, 2022, pp. 112--124.

\bibitem{harkous2022hark}
H.~Harkous, S.~T. Peddinti, R.~Khandelwal, A.~Srivastava, and N.~Taft, ``Hark: A deep learning system for navigating privacy feedback at scale,'' in \emph{2022 IEEE Symposium on Security and Privacy (SP)}.\hskip 1em plus 0.5em minus 0.4em\relax IEEE, 2022, pp. 2469--2486.

\bibitem{ebrahimi2022unsupervised}
F.~Ebrahimi and A.~Mahmoud, ``Unsupervised summarization of privacy concerns in mobile application reviews,'' in \emph{Proc. of the 37th IEEE/ACM International Conference on Automated Software Engineering}, 2022, pp. 1--12.

\bibitem{nguyen2019short}
D.~C. Nguyen, E.~Derr, M.~Backes, and S.~Bugiel, ``Short text, large effect: Measuring the impact of user reviews on android app security \& privacy,'' in \emph{2019 IEEE symposium on Security and Privacy (SP)}.\hskip 1em plus 0.5em minus 0.4em\relax IEEE, 2019, pp. 555--569.

\bibitem{mukherjee2020empirical}
D.~Mukherjee, A.~Ahmadi, M.~V. Pour, and J.~Reardon, ``An empirical study on user reviews targeting mobile apps’ security \& privacy,'' \emph{arXiv preprint arXiv:2010.06371}, 2020.

\bibitem{dong2022privacy}
Z.~Dong, L.~Wang, H.~Xie, G.~Xu, and H.~Wang, ``Privacy analysis of period tracking mobile apps in the post-roe v. wade era,'' in \emph{Proceedings of the 37th IEEE/ACM International Conference on Automated Software Engineering}, 2022, pp. 1--6.

\bibitem{fazelnia2024lessons}
M.~Fazelnia, V.~Koscinski, S.~Herzog, and M.~Mirakhorli, ``Lessons from the use of natural language inference (nli) in requirements engineering tasks,'' in \emph{2024 IEEE 32nd International Requirements Engineering Conference (RE)}.\hskip 1em plus 0.5em minus 0.4em\relax IEEE, 2024, pp. 103--115.

\bibitem{demszky2018transforming}
D.~Demszky \emph{et~al.}, ``Transforming question answering datasets into natural language inference datasets,'' \emph{arXiv preprint arXiv:1809.02922}, 2018.

\bibitem{stasaski2022semantic}
K.~Stasaski \emph{et~al.}, ``Semantic diversity in dialogue with natural language inference,'' in \emph{Proceedings of the 2022 Conference of the North American Chapter of the Association for Computational Linguistics: Human Language Technologies}.\hskip 1em plus 0.5em minus 0.4em\relax Association for Computational Linguistics, Jul. 2022, pp. 85--98.

\bibitem{chae2023large}
Y.~Chae \emph{et~al.}, ``Large language models for text classification: From zero-shot learning to fine-tuning,'' \emph{Open Science Foundation}, vol.~10, 2023.

\bibitem{maccartney2009extended}
B.~MacCartney \emph{et~al.}, ``An extended model of natural logic,'' in \emph{Proc. of the eight international conference on computational semantics}, 2009, pp. 140--156.

\bibitem{iwaya2023privacy}
L.~H. Iwaya \emph{et~al.}, ``On the privacy of mental health apps: An empirical investigation and its implications for app development,'' \emph{Empirical Software Engineering}, vol.~28, no.~1, p.~2, 2023.

\bibitem{devlin2019bert}
J.~Devlin, M.-W. Chang, K.~Lee, and K.~Toutanova, ``Bert: Pre-training of deep bidirectional transformers for language understanding,'' in \emph{Proceedings of the 2019 conference of the North American chapter of the association for computational linguistics: human language technologies, volume 1 (long and short papers)}, 2019, pp. 4171--4186.

\bibitem{liu2019roberta}
Y.~Liu \emph{et~al.}, ``Roberta: A robustly optimized bert pretraining approach,'' \emph{arXiv preprint arXiv:1907.11692}, 2019.

\bibitem{he2021deberta}
\BIBentryALTinterwordspacing
P.~He \emph{et~al.}, ``Deberta: Decoding-enhanced bert with disentangled attention,'' in \emph{2021 International Conference on Learning Representations}, May 2021, under review. [Online]. Available: \url{https://www.microsoft.com/en-us/research/publication/deberta-decoding-enhanced-bert-with-disentangled-attention-2/}
\BIBentrySTDinterwordspacing

\bibitem{he2021debertav3}
{He, Pengcheng} \emph{et~al.}, ``Debertav3: Improving deberta using electra-style pre-training with gradient-disentangled embedding sharing,'' \emph{arXiv preprint arXiv:2111.09543}, 2021.

\bibitem{wolf2019huggingface}
T.~Wolf \emph{et~al.}, ``Transformers: State-of-the-art natural language processing,'' in \emph{Proceedings of the 2020 Conference on Empirical Methods in Natural Language Processing: System Demonstrations}.\hskip 1em plus 0.5em minus 0.4em\relax Online: Association for Computational Linguistics, Oct. 2020, pp. 38--45.

\bibitem{williams2017broad}
A.~Williams \emph{et~al.}, ``A broad-coverage challenge corpus for sentence understanding through inference,'' in \emph{Proceedings of the 2018 Conference of the North {A}merican Chapter of the Association for Computational Linguistics: Human Language Technologies, Volume 1 (Long Papers)}.\hskip 1em plus 0.5em minus 0.4em\relax Association for Computational Linguistics, Jun. 2018, pp. 1112--1122.

\bibitem{nie2019adversarial}
Y.~Nie \emph{et~al.}, ``Adversarial {NLI}: A new benchmark for natural language understanding,'' in \emph{Proceedings of the 58th Annual Meeting of the Association for Computational Linguistics}.\hskip 1em plus 0.5em minus 0.4em\relax Association for Computational Linguistics, Jul. 2020, pp. 4885--4901.

\bibitem{thorne2019fever2}
J.~Thorne \emph{et~al.}, ``The fever2. 0 shared task,'' in \emph{Proc. of the second workshop on Fact Extraction and VERification}, 2019, pp. 1--6.

\bibitem{duvsek2020evaluating}
O.~Du{\v{s}}ek \emph{et~al.}, ``Evaluating semantic accuracy of data-to-text generation with natural language inference,'' in \emph{Proceedings of the 13th International Conference on Natural Language Generation}.\hskip 1em plus 0.5em minus 0.4em\relax Association for Computational Linguistics, Dec. 2020, pp. 131--137.

\bibitem{chen2023unleashing}
B.~Chen \emph{et~al.}, ``Unleashing the potential of prompt engineering for large language models,'' \emph{Patterns}, 2025.

\bibitem{touvron2023llama}
H.~Touvron \emph{et~al.}, ``Llama: Open and efficient foundation language models,'' \emph{arXiv preprint arXiv:2302.13971}, 2023.

\bibitem{zhang2023revisiting}
T.~Zhang \emph{et~al.}, ``Revisiting sentiment analysis for software engineering in the era of large language models,'' \emph{ACM Trans. Softw. Eng. Methodol.}, vol.~34, no.~3, Feb. 2025.

\bibitem{ouyang2023llm}
S.~Ouyang \emph{et~al.}, ``An empirical study of the non-determinism of chatgpt in code generation,'' \emph{ACM Trans. Softw. Eng. Methodol.}, vol.~34, no.~2, Jan. 2025.

\bibitem{ye2023comprehensive}
J.~Ye \emph{et~al.}, ``A comprehensive capability analysis of gpt-3 and gpt-3.5 series models,'' \emph{arXiv preprint arXiv:2303.10420}, 2023.

\bibitem{hurst2024gpt}
\BIBentryALTinterwordspacing
OpenAI, ``Gpt-4o mini: Advancing cost-efficient intelligence,'' Jul 2024. [Online]. Available: \url{https://openai.com/index/gpt-4o-mini-advancing-cost-efficient-intelligence}
\BIBentrySTDinterwordspacing

\bibitem{wang2021want}
S.~Wang \emph{et~al.}, ``Want to reduce labeling cost? {GPT}-3 can help,'' in \emph{Findings of the Association for Computational Linguistics: EMNLP 2021}.\hskip 1em plus 0.5em minus 0.4em\relax Association for Computational Linguistics, Nov. 2021, pp. 4195--4205.

\bibitem{sebastiani2002machine}
F.~Sebastiani, ``Machine learning in automated text categorization,'' \emph{ACM computing surveys (CSUR)}, vol.~34, no.~1, pp. 1--47, 2002.

\bibitem{cohen1968weighted}
J.~Cohen, ``Weighted kappa: Nominal scale agreement provision for scaled disagreement or partial credit.'' \emph{Psychological bulletin}, vol.~70, no.~4, p. 213, 1968.

\bibitem{joachims1998text}
T.~Joachims, ``Text categorization with support vector machines: Learning with many relevant features,'' in \emph{European conference on machine learning}.\hskip 1em plus 0.5em minus 0.4em\relax Springer, 1998, pp. 137--142.

\bibitem{basu2003support}
A.~Basu \emph{et~al.}, ``Support vector machines for text categorization,'' in \emph{36th Annual Hawaii International Conference on System Sciences, 2003. Proceedings of the}.\hskip 1em plus 0.5em minus 0.4em\relax IEEE, 2003, pp. 7--pp.

\bibitem{solove2005taxonomy}
D.~J. Solove, ``A taxonomy of privacy,'' \emph{U. Pa. l. Rev.}, vol. 154, p. 477, 2005.

\bibitem{wang2009privacy}
Y.~Wang, ``Privacy-enhancing technologies,'' in \emph{Handbook of research on social and organizational liabilities in information security}.\hskip 1em plus 0.5em minus 0.4em\relax IGI Global, 2009, pp. 203--227.

\bibitem{viera2005understanding}
A.~J. Viera \emph{et~al.}, ``Understanding interobserver agreement: the kappa statistic,'' \emph{Fam med}, vol.~37, no.~5, pp. 360--363, 2005.

\bibitem{zhu2023can}
Y.~Zhu \emph{et~al.}, ``Exploring the capability of chatgpt to reproduce human labels for social computing tasks,'' in \emph{International Conference on Advances in Social Networks Analysis and Mining}.\hskip 1em plus 0.5em minus 0.4em\relax Springer, 2024, pp. 13--22.

\bibitem{moller2023parrot}
A.~G. M{\o}ller \emph{et~al.}, ``The parrot dilemma: Human-labeled vs. {LLM}-augmented data in classification tasks,'' in \emph{Proceedings of the 18th Conference of the European Chapter of the Association for Computational Linguistics (Volume 2: Short Papers)}.\hskip 1em plus 0.5em minus 0.4em\relax Association for Computational Linguistics, Mar. 2024, pp. 179--192.

\bibitem{li2022narratives}
Z.~S. Li, M.~Sihag, N.~N. Arony, J.~B. Junior, T.~Phan, N.~Ernst, and D.~Damian, ``Narratives: the unforeseen influencer of privacy concerns,'' in \emph{2022 IEEE 30th International Requirements Engineering Conference (RE)}.\hskip 1em plus 0.5em minus 0.4em\relax IEEE, 2022, pp. 127--139.

\bibitem{dubey2024llama}
A.~Dubey \emph{et~al.}, ``The llama 3 herd of models,'' \emph{arXiv preprint arXiv:2407.21783}, 2024.

\bibitem{almazrouei2023falcon}
\BIBentryALTinterwordspacing
E.~Almazrouei \emph{et~al.}, ``Introducing the technology innovation institute’s falcon 3 making advanced ai accessible and available to everyone, everywhere,'' Nov 2023. [Online]. Available: \url{https://falconllm.tii.ae/falcon-models.html}
\BIBentrySTDinterwordspacing

\bibitem{jiang2023mistral}
\BIBentryALTinterwordspacing
A.~Q. Jiang \emph{et~al.}, ``Mistral 7b,'' Sep 2023. [Online]. Available: \url{https://mistral.ai/news/announcing-mistral-7b}
\BIBentrySTDinterwordspacing

\bibitem{alomar2021finding}
E.~A. AlOmar \emph{et~al.}, ``Finding the needle in a haystack: On the automatic identification of accessibility user reviews,'' in \emph{Proc. of the 2021 CHI conference on human factors in computing systems}, 2021, pp. 1--15.

\bibitem{hou2023large}
X.~Hou \emph{et~al.}, ``Large language models for software engineering: A systematic literature review,'' \emph{ACM Trans. Softw. Eng. Methodol.}, vol.~33, no.~8, Dec. 2024.

\bibitem{colavito2024leveraging}
G.~Colavito, F.~Lanubile, N.~Novielli, and L.~Quaranta, ``Leveraging gpt-like llms to automate issue labeling,'' in \emph{2024 IEEE/ACM 21st International Conference on Mining Software Repositories (MSR)}.\hskip 1em plus 0.5em minus 0.4em\relax IEEE, 2024, pp. 469--480.

\bibitem{arora2024advancing}
C.~Arora, J.~Grundy, and M.~Abdelrazek, ``Advancing requirements engineering through generative ai: Assessing the role of llms,'' in \emph{Generative AI for Effective Software Development}.\hskip 1em plus 0.5em minus 0.4em\relax Springer, 2024, pp. 129--148.

\bibitem{bencheikh2023exploring}
L.~Bencheikh and N.~H{\"o}glund, ``Exploring the efficacy of chatgpt in generating requirements: An experimental study,'' 2023.

\bibitem{ronanki2023investigating}
K.~Ronanki, C.~Berger, and J.~Horkoff, ``Investigating chatgpt’s potential to assist in requirements elicitation processes,'' in \emph{2023 49th Euromicro Conference on Software Engineering and Advanced Applications (SEAA)}.\hskip 1em plus 0.5em minus 0.4em\relax IEEE, 2023, pp. 354--361.

\bibitem{white2024chatgpt}
J.~White, S.~Hays, Q.~Fu, J.~Spencer-Smith, and D.~C. Schmidt, ``Chatgpt prompt patterns for improving code quality, refactoring, requirements elicitation, and software design,'' in \emph{Generative ai for effective software development}.\hskip 1em plus 0.5em minus 0.4em\relax Springer, 2024, pp. 71--108.

\bibitem{rasheed2024autonomous}
Z.~Rasheed \emph{et~al.}, ``Autonomous agents in software development: A vision paper,'' in \emph{International Conference on Agile Software Development}.\hskip 1em plus 0.5em minus 0.4em\relax Springer Nature Switzerland Cham, 2024, pp. 15--23.

\bibitem{sridhara2023chatgpt}
G.~Sridhara \emph{et~al.}, ``Chatgpt: A study on its utility for ubiquitous software engineering tasks,'' \emph{arXiv preprint arXiv:2305.16837}, 2023.

\bibitem{marques2024using}
N.~Marques, R.~R. Silva, and J.~Bernardino, ``Using chatgpt in software requirements engineering: A comprehensive review,'' vol.~16, no.~6, p. 180, 2024.

\end{thebibliography}

\end{document}